# Application of Clustering Methods to Anomaly Detection in Fibrous Media

**Denis Dresvyanskiy[1], Tatiana Karaseva[1], Sergei Mitrofanov[1], Claudia Redenbach[2], Stefanie Schwaar[2], Vitalii Makogin[3], Evgeny Spodarev[3]**

[1]Reshetnev Siberian State University of Science and Technology
31, Krasnoyarsky Rabochy Av., Krasnoyarsk, 660037, Russian Federation
[2]Technische Universität Kaiserslautern, Fachbereich Mathematik, Postfach 3049, 67653 Kaiserslautern, Germany
[3]Institut für Stochastik, Universität Ulm, D-89069 Ulm, Germany

E-mail: tatyanakarasewa@yandex.ru

**Abstract**. The paper considers the problem of anomaly detection in 3D images of fibre materials. The spatial Stochastic Expectation Maximisation algorithm and Adaptive Weights Clustering are applied to solve this problem. The initial 3D grey scale image was divided into small cubes subject to clustering. For each cube clustering attributes values were calculated: mean local direction and directional entropy. Clustering is conducted according to the given attributes. The proposed methods are tested on the simulated images and on real fibre materials.

**Introduction.**
Nowadays, there exists a large amount of novel materials with interesting physical properties. For instance, reinforcement of polymers with fibres significantly increases the mechanical properties of the materials. The materials' performance is determined mostly by their composition as well as by allocation and directions of the reinforcing fibres.

Due to the production process, an anomaly region may be formed in the material. We define it as an area where the distribution of fibre directions differs from the remaining material. To keep the stated material's properties, it is necessary to identify regions with untypical fibre distribution. For this purpose, high-resolution microcomputer tomography reaching a level of microns [1, 2] is used to observe the fibre system in a composite sample, cf. Figure 1 (right).

Our main task is then to find the areas with anomalous directional properties of fibres in the 3D image. This is done by means of cluster analysis dividing the whole image volume into two clusters: the smaller "anomaly" region and the bigger "normal" material.

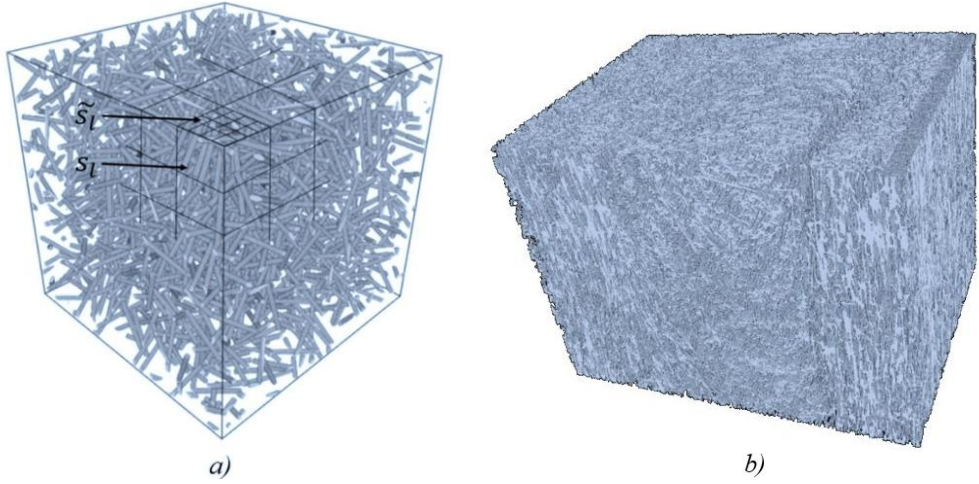

**Figure 1.** 3D images a) RSA generated fibres, size in voxels: 200x200x300, b) real glass-fibre reinforced data, size in voxels: 1500x1000x1000. Courtesy of the Composite Materials Institute (IVW), Kaiserslautern

The paper consists of three parts. In the first part, we introduce clustering criteria. In the second part, we consider two clustering methods: the spatial Stochastic Expectation Maximisation (SEM) [3, 4] and the Adaptive Weights Clustering (AWC) [5], and also test them on simulated data. In the third part, we apply our algorithms to real fibre data and compare their performance.

**1. Clustering criteria.**
Let us have a 3D image of the fibre material. It is assumed that it may contain a region where the local distribution of fibres does not coincide with their distribution throughout the remaining material volume. It is necessary to determine the criteria according to which such an anomaly region can be located automatically. In this study, the criteria are chosen to be the entropy of the local directional distribution [6, 7] and the mean of the local fibre directions.

The necessary 3D image preprocessing prior to our clustering was performed by the Modular Algorithms for Volume Images (MAVI) software by using the method based on the Hessian matrix presented in [8]. Using MAVI, the local directions $(x, y, z)^T$ were obtained as the mean in a small neighborhood (small cubes $\widetilde{s}_l$ in Figure 1). Thus, one calculated local direction corresponds to a cube of voxels which is done for the sake of computational efficiency.

*1.1. Mean of local direction of fibres.*

This criterion is calculated as the coordinate wise mean value of local directions $(x, y, z)^T$ in the scanning window $s_l$. Let the local directions $(x_i, y_i, z_i), i = 1 \ldots N$, be given in $s_l$. Then the mean vector (X, Y, Z) of local directions is defined as:

$$X = \frac{1}{N}\sum_{i=1}^{N} x_i \,;\, Y = \frac{1}{N}\sum_{i=1}^{N} y_i \,;\, Z = \frac{1}{N}\sum_{i=1}^{N} z_i.$$

This criterion has three components, so clustering according to the mean of local fibre directions is carried out in the space $R^3$.

*1.2. Entropy of local directional distribution.*

The differential (Shannon) entropy is represented as a functional:

$$H = H(f) = -\int_{S^2} f(x) \ln f(x) \, dx, \tag{1}$$

where $S^2$ is the 2-dimensional unit sphere in the Euclidean space with geodesic metric $\rho$, and $f(x)$ is directional distribution density with respect to the surface area measure $dx$.

*1.2.1 Nearest neighbour entropy estimator.*
The following idea for estimation of the entropy goes back to Dobrushin [9]. Assume we have a sample of random direction vectors $X_1, \ldots, X_N, N \geq 2$ on the sphere with their geographic reference points lying in $s_l$. We calculate $\rho_i = \min\{\rho(X_i, X_j), j \in \{1, \ldots, N\setminus\{i\}\}$ and define:

$$\bar{\rho} = \left\{\prod_{i=1}^{N} \rho_i\right\}.$$

The nearest neighbour entropy estimator is given by
$$\hat{H} = 2\ln\bar{\rho} + \ln c_1 + \gamma + \ln(N-1), \qquad (2)$$

where $\gamma \approx 0{,}5772$ is the Euler's constant and $c_1 \approx 1{,}1447$.

Here we consider clustering of the image volume according to the above local entropy estimate and vectors of the mean of local directions.

## 2. Clustering.

While calculating the entropy of local directions in fibre materials, it turned out that the entropy histogram in a homogeneous fibrous medium has one mode, and two modes if a large enough anomaly is present cf. [4]. Hence, it is reasonable to consider the problem of separation of modes in a mixture of two normal distributions in order to perform the desired clustering into an anomaly and a normal material volume. In this investigation, the cluster objects are windows $s_l$, in which the values of the clustering criteria $(\hat{H}, (X, Y, Z)^T)$ were calculated as mentioned above. To solve this clustering problem, the spatial modification of SEM [4] and AWC algorithms are used.

### 2.1. RSA simulated image data

Now we illustrate the use of spatial SEM and AWC methods on a simulated 3D fibre image, compare Figure 2. We choose a layered random sequential absorption (RSA, see [10, 11]) model. That is, the fibres are added randomly to the existing material, so that they cannot intersect each other. The clustering results for both methods are shown in Figures 3-5. There, the red dots refer to the anomaly region, blue ones are homogeneous material, and the green dots represent the third cluster (artefact) found by AWC.

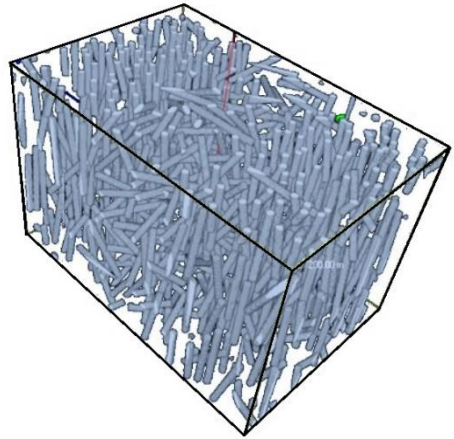

**Figure 2.** RSA data

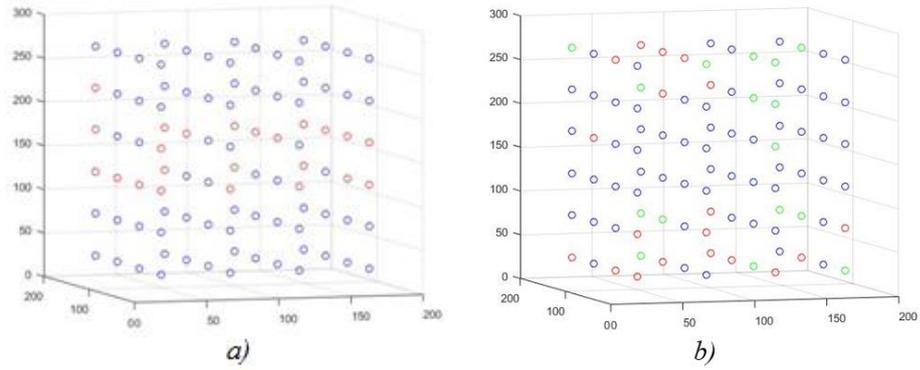

**Figure 3.** Clustering RSA data with entropy attribute. Algorithms: a) spatial SEM, b) AWC

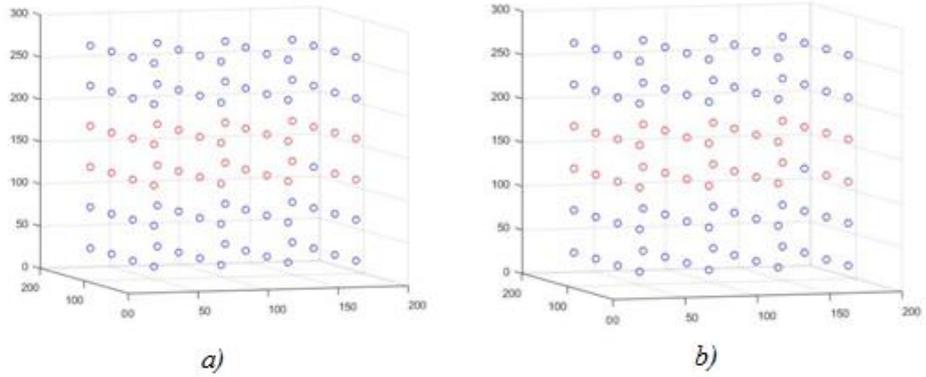

**Figure 4.** Clustering RSA data with mean of local direction attribute. Algorithms: a) spatial SEM, b) AWC

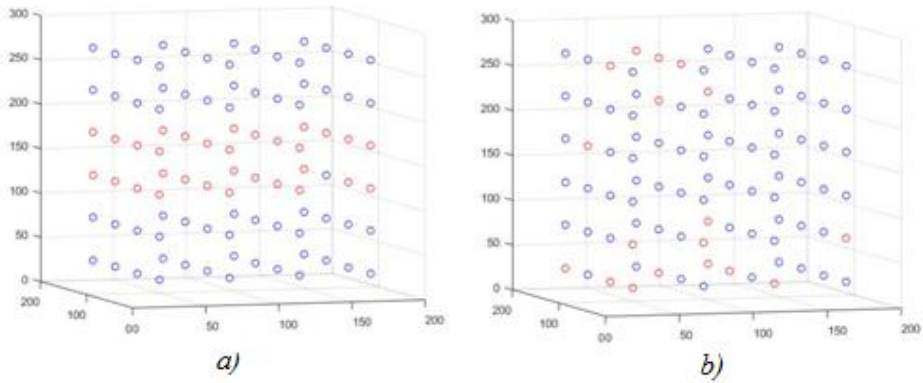

**Figure 5.** Results Clustering RSA data with mean local direction and entropy attributes. Algorithms: a) spatial SEM, b) AWC

In Figure 4, the mean of local directions is used as criterion and both algorithms detect the anomaly up to one cube $s_l$. On the other hand, when entropy or mean of local directions with entropy are applied as a criterion for clustering, the AWC algorithm finds 3 clusters (instead of two) and has a large error.

Thus, the obtained results demonstrate a better efficiency of spatial SEM in comparison to AWC in performing the anomaly detection for these data. The initial number of classes in the spatial SEM is the only parameter to be tuned. It is chosen to be 10. The AWC's only setting is parameter $\lambda$; its values are given in Table 1.

**Table 1.** Values of parameter λ for the AWC algorithm used in our calculations

| Clustering attribute | Entropy | Mean of local direction | Mean of local direction and entropy |
|---|---|---|---|
| RSA data | 9.2 | 10 | 0.2 |
| Real fibre data | 19.27 | 4.27 | 1.21 |

*2.2 Real fibre materials.*

Appling our approach to detect anomalies in a 3D image of a real glass-fibre reinforced composite material shown in Figure 1 b). The results of the clustering are presented in Figures 6-8. The color encoding is as follows: red stands for anomaly, blue for the homogeneous material, and all other colors (green, yellow, etc.) indicate different artefact clusters found by AWC.

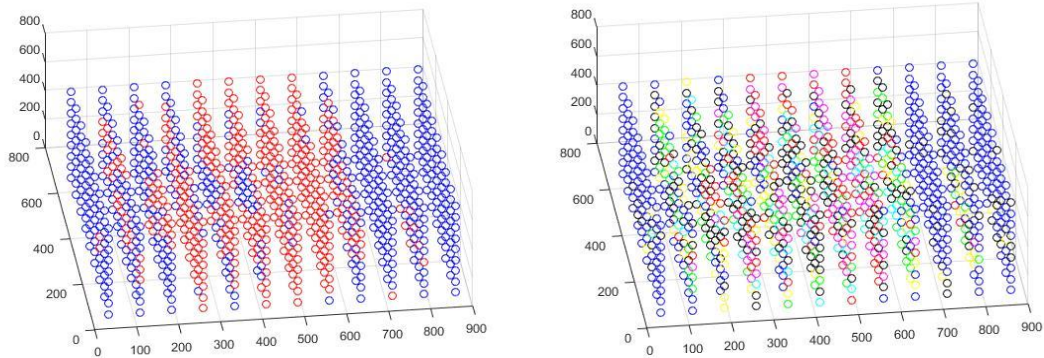

**Figure 6.** Clustering real fibre data with entropy attribute. Algorithms: a) spatial SEM, b) AWC

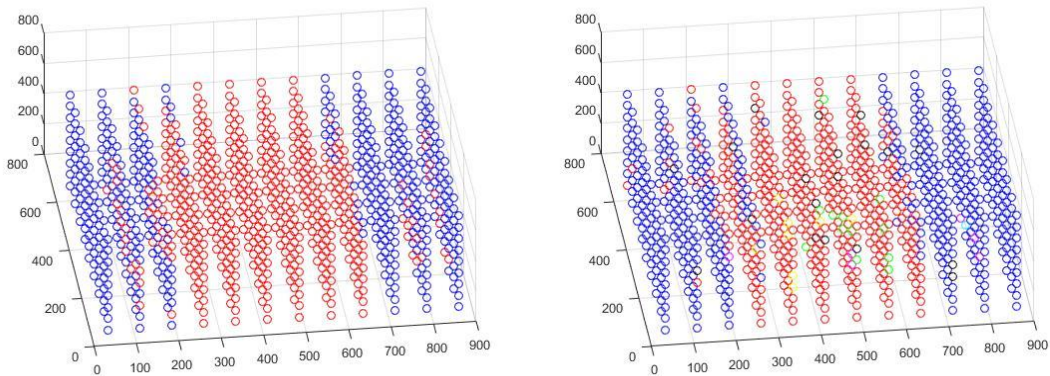

**Figure 7.** Clustering real fibre data with mean of local direction attribute. Algorithms: a) spatial SEM, b) AWC

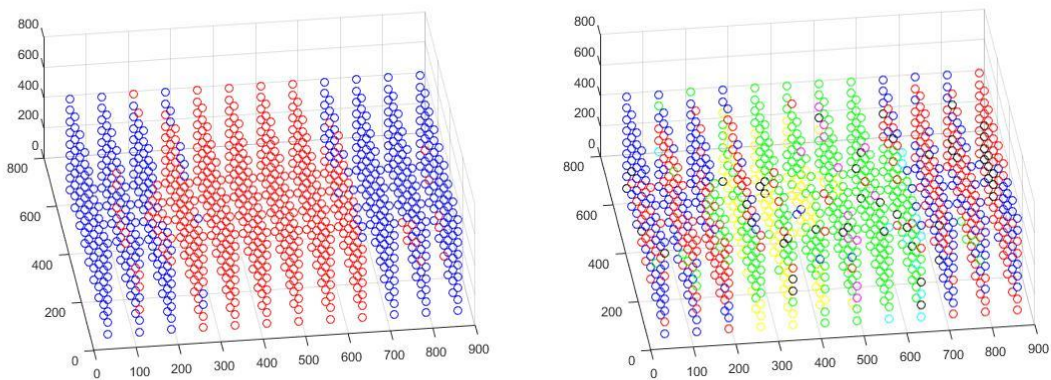

**Figure 8.** Clustering real fibre data with mean of local direction and entropy attributes. Algorithms: a) spatial SEM, b) AWC

The estimated anomaly region is also located in the center of the considered part of the real glass fibre-reinforced composite material. As one can see in the figures, the SEM algorithm accurately determines an anomaly region. Moreover the application of various clustering criteria helps to detect various types of anomaly, that is: the entropy detects a vortex-like anomaly, whereas changes in the average direction of fibres are best detected by the attribute of the mean of local direction. The AWC algorithm showed low efficiency with real image data. For any clustering criteria, we found at least 17 clusters which cannot be easily subdivided into two clear groups pertaining to anomaly and to the normal material, while the SEM algorithm is designed to find only these two classes. Therefore, the spatial SEM algorithm is preferable for this task.

**Conclusion.**
As a result of this research, a software system was developed to identify an anomaly region in 3-dimensional grayscale images of fibre materials. The software system consists of an algorithmic complex. It involves the computation of local directions in MAVI, the calculation of clustering criteria and the spatial SEM algorithm.

**Acknowledgement.**
Stefanie Schwaar is grateful for funding within the DFG research training group 1932 "Stochastic Models for Innovations in the Engineering Sciences".